\documentclass[smallextended]{svjour3}       
\smartqed  

\RequirePackage[OT1]{fontenc}

\usepackage{amssymb,mathrsfs,lineno,adjustbox,tabularx,ragged2e,slashbox,float, physics}

\usepackage{multirow}
\usepackage{geometry}

\usepackage{amsmath}
\usepackage{algorithm}
\usepackage{algorithmic}
\usepackage{parselines}
\usepackage{epsfig}
\usepackage[nottoc]{tocbibind}
\usepackage[colorlinks,citecolor=blue,urlcolor=blue,filecolor=blue,backref=page]{hyperref}
\usepackage{graphicx}
\usepackage{graphics}
\usepackage{subfigure}
\usepackage{verbatim}
\usepackage{float}
\usepackage{mathtools}
\usepackage{enumerate}
\usepackage{natbib}
\usepackage{url}

\usepackage{mathptmx}      
\begin{document}

\title{Bayesian analysis of absolute continuous Marshall-Olkin bivariate Pareto distribution with location and scale parameters} 


\titlerunning{EM-BVPA}        
\author{Biplab Paul, Arabin Kumar Dey and Sanku Dey\\
}

           

\institute{B. Paul \at
             Department of Mathematics,\\ 
             IIT Guwahati,\\   
             Guwahati, India\\
             \email{biplab.paul@iitg.ac.in}  \\ 
\\
            A. K. Dey \at
              Department of Mathematics, \\
              IIT Guwahati,\\
              Guwahati, India\\
              Assam\\
              Tel.: +91361-258-4620\\
              \email{arabin@iitg.ac.in}\\ 
\\
  	    Sanku Dey \at
	      Department of Statistics,\\	
              St. Anthony's College, Shillong.\\	            
}



\maketitle

\begin{abstract}

  This paper provides two different novel approaches of slice sampling to estimate the parameters of absolute continuous Marshall-Olkin bivariate Pareto distribution with location and scale parameters. We carry out the bayesian analysis taking gamma prior for shape and scale parameters and truncated normal for location parameters.  Credible intervals and coverage probabilities are also provided for all methods.  A real-life data analysis is shown for illustrative purpose.    

\end{abstract}


\section{Introduction}
\label{intro}

 In this paper we consider bayesian analysis of the parameters of absolute continuous Marshall-Olkin bivariate Pareto distribution with location and scale parameters through slice sampling.  Usual slice sampling won't work in monte carlo set up as posterior function in this case becomes discontinuous with respect to location and scale parameters.  We propose two novel approaches in implementation of slice sampling to perform the bayesian analysis. 

  Recently three parameter bayesian analysis of absolute continuous version of Marshall-Olkin bivariate Pareto distribution is studied by \cite{DeyPaul:2017}.  This absolute continuous version of Marshall-Olkin bivariate Pareto distribution has marginals which are not type-II univariate Pareto distributions.  We use the notation BB-BVPA for absolute continuous Marshall-Olkin bivariate Pareto.  This form of Marshall-Olkin bivariate Pareto is similar to absolute continuous bivariate exponential distribution as proposed by \cite{BlockBasu:1974}.  A variety of bivariate (multivariate) extensions of bivariate Pareto distribution also have been studied in the literature.  These include the distributions described in the following works : \cite{SankaranKundu:2014}, \cite{Yeh:2000}, \cite{Yeh:2004}, \cite{AsimitFurmanVernic:2010}, \cite{AsimitFurmanVernic:2016}.  Finding efficient technique to estimate the parameters of BB-BVPA, particularly in the presence of location and scale was a major challenge in last few decades.  Parameter estimation by EM algorithm for BB-BVPA is also available in a recent work by \cite{DeyKundu:2017}.  There is no work available for seven parameter bayesian analysis on BB-BVPA.

   Finding confidence interval for location and scale parameters for BB-BVPA is a difficult problem in frequentist set up.  Bayesian credible intervals can easily solve this problem.  Sometimes Bayes estimators exist when MLEs do not exist.  Bayesian estimators may work reasonably well with suitable choice of prior even when MLE's performance is extremely poor.  Therefore working in bayesian set up with such a complicated distribution has its own advantages.  In this paper we use gamma prior for shape and scale parameters.  However for location parameters we use truncated normal as prior. 

 Rest of the paper is organized as follows.  Section 2 shows formulation of absolute continuous Marshall-Olkin bivariate Pareto distribution. Bayesian analysis through slice sampling is described in Section 3.  Section 4 deals with the construction of credible interval.  Numerical results are discussed in section 5.  Data analysis is shown in section 6.  We conclude the paper in section 7.        

\section{Marshall-Olkin bivariate Pareto Distribution:}

  Let $U_0, U_1$ and $U_2$ are mutually  independent random variable where $U_0 \sim PA(II)(0, 1, \alpha_0)$, $U_1 \sim PA(II)(\mu_1, \sigma_1,\alpha_1)$ and  $U_2 \sim PA(II)(\mu_2, \sigma_2, \alpha_2)$. We define $X_1 = \min\{\mu_1 + \sigma_1U_0, U_1\}$ and $X_2 = \min\{\mu_2 + \sigma_2U_0, U_1\}$, then the joint distribution  of $(X_1, X_2)$ is called the \textbf{Marshall-Olkin bivariate Pareto (MOBVPA) distribution} or singular bivariate Pareto distribution. The joint survival function of $(X_1, X_2)$ can be written for $z = \max\{\frac{x_1 - \mu_1}{\sigma_1}, \frac{x_2 - \mu_2}{\sigma_2}\}$ as;

\begin{align*}
S(x_1, x_2)&= (1 + z)^{-\alpha_0}\Big(1 + \frac{x_1 - \mu_1}{\sigma_1}\Big)^{-\alpha_1}\Big(1 + \frac{x_2 - \mu_2}{\sigma_2}\Big)^{-\alpha_2}\\&=
\begin{cases}
S_1(x_1, x_2), \quad \text{if $\frac{x_1 - \mu_1}{\sigma_1}$ \textless $\frac{x_2 - \mu_2}{\sigma_2}$}
\\
S_2(x_1, x_2),\quad \text{if $\frac{x_1 - \mu_1}{\sigma_1}$ \textgreater $\frac{x_2 - \mu_2}{\sigma_2}$}\\
S_{0}(x), \quad \text{if $\frac{x_1 - \mu_1}{\sigma_1}$ = $\frac{x_2 - \mu_2}{\sigma_2}$} \quad \text{and}\quad  x_1=x_2=x
\end{cases}
\end{align*}
where 
\begin{align*}
S_1(x_1, x_2)&=  \Big(1 + \frac{x_2 - \mu_2}{\sigma_2}\Big)^{- \alpha_0 - \alpha_2} \Big(1 + \frac{x_1 - \mu_1}{\sigma_1}\Big)^{-\alpha_1}\\
S_2(x_1, x_2)&=  \Big(1 + \frac{x_2 - \mu_2}{\sigma_2}\Big)^{-\alpha_2}\Big(1 + \frac{x_1 - \mu_1}{\sigma_1}\Big)^{- \alpha_0 - \alpha_1}\\
S_0(x)&=\Big(1 + \frac{x - \mu_1}{\sigma_1}\Big)^{-\alpha_0 - \alpha_1 - \alpha_2}
\end{align*}

so it's pdf can be written as 
\begin{align*}
f(x_1, x_2)&=
\begin{cases}
f_1(x_1, x_2),\quad \text{if $\frac{x_1 - \mu_1}{\sigma_1}$ \textless $\frac{x_2 - \mu_2}{\sigma_2}$}
\\
f_2(x_1, x_2),\quad \text{if $\frac{x_1 - \mu_1}{\sigma_1}$ \textgreater $\frac{x_2 - \mu_2}{\sigma_2}$}\\
f_{0}(x), \quad \text{if $\frac{x_1 - \mu_1}{\sigma_1}$ = $\frac{x_2 - \mu_2}{\sigma_2}$} \quad \text{and}\quad  x_1=x_2=x
\end{cases}
\end{align*}
where 
\begin{align*}
f_1(x_, x_2)&=\frac{\alpha_1 (\alpha_0 + \alpha_2)}{\sigma_1 \sigma_2}\Big(1 + \frac{x_2 - \mu_2}{\sigma_2}\Big)^{- \alpha_0 - \alpha_2 - 1}\Big(1 + \frac{x_1 - \mu_1}{\sigma_1}\Big)^{- \alpha_1 - 1}\\
f_2(x_1, x_2)&=\frac{\alpha_2 (\alpha_0 + \alpha_1)}{\sigma_1 \sigma_2}\Big(1 + \frac{x_2 - \mu_2}{\sigma_2}\Big)^{-\alpha_2 - 1}\Big(1 + \frac{x_1 - \mu_1}{\sigma_1}\Big)^{- \alpha_0 - \alpha_1 - 1} \\
f_0(x)&= \frac{\alpha_0}{\sigma_1}\Big(1 + \frac{x - \mu_1}{\sigma_1}\Big)^{- \alpha_0 - \alpha_1 - \alpha_2 - 1}
\end{align*}
We denote this distribution as $MOBVPA(\mu_1, \mu_2, \sigma_1, \sigma_2, \alpha_0, \alpha_1, \alpha_2)$.

\subsection{Block Basu bivariate Pareto Distribution}

  We know that joint survival function of $(X_1, X_2)$ can be written as a mixture of an absolute continuous part and a singular part as follows;
\begin{equation*}
S(x_1, x_2) = p S_a(x_1, x_2) + (1 - p) S_s(x_1, x_2)
\end{equation*}
where $S_a(x_1, x_2)$  is the absolute continuous part and $S_s(x_1, x_2)$ is the singular part.  Also $ p = \frac{\alpha_1 + \alpha_2}{\alpha_0 + \alpha_1 + \alpha_2}$. Note that for $z = \max\{\frac{x_1 - \mu_1}{\sigma_1}, \frac{x_2 - \mu_2}{\sigma_2}\}$,

\begin{equation*}
S_s(x_1, x_2)=(1 + z)^{-\alpha_0 - \alpha_1 - \alpha_2}
\end{equation*}

  We define $S_a(x_1, x_2)$ as the joint survival function of \textbf{Block Basu bivariate Pareto (BBBVPA) distribution} or absolute continuous bivariate pareto distribution.  It can then be expressed as 

\begin{align*}
S_a(x_1, x_2)&=
\begin{cases}
S_{a1}(x_1, x_2), \quad \text{if $\frac{x_1 - \mu_1}{\sigma_1}$ \textless $\frac{x_2 - \mu_2}{\sigma_2}$}
\\
S_{a2}(x_1, x_2), \quad \text{if $\frac{x_1 - \mu_1}{\sigma_1}$ \textgreater $\frac{x_2 - \mu_2}{\sigma_2}$} 
\end{cases}
\end{align*} 
where 
\begin{align*}
S_{a1}(x_1, x_2) = \frac{\alpha_0 + \alpha_1 + \alpha_2}{\alpha_1 + \alpha_2}\Big(1 + \frac{x_2 - \mu_2}{\sigma_2}\Big)^{-\alpha_0 - \alpha_2}\Big(1 + \frac{x_1 - \mu_1}{\sigma_1}\Big)^{-\alpha_1} \\- \frac{\alpha_0}{\alpha_1 + \alpha_2}\Big(1 + \frac{x_2 - \mu_2}{\sigma_2}\Big)^{-\alpha_0 - \alpha_1 - \alpha_2}\\
S_{a2}(x_1, x_2)=  \frac{\alpha_0 + \alpha_1 + \alpha_2}{\alpha_1 + \alpha_2}\Big(1 + \frac{x_2 - \mu_2}{\sigma_2}\Big)^{-\alpha_2}\Big(1 + \frac{x_1 - \mu_1}{\sigma_1}\Big)^{-\alpha_0 - \alpha_1} \\- \frac{\alpha_0}{\alpha_1 + \alpha_2}\Big(1 + \frac{x_1 - \mu_1}{\sigma_1}\Big)^{-\alpha_0 - \alpha_1 - \alpha_2}
\end{align*}
so it's pdf can be written as 
\begin{align*}
f_{a}(x_1, x_2)&=
\begin{cases}
\dfrac{1}{p}f_1(x_1, x_2),\quad \text{if $\frac{x_1 - \mu_1}{\sigma_1}$ \textless $\frac{x_2 - \mu_2}{\sigma_2}$}
\\
\frac{1}{p}f_2(x_1, x_2),\quad \text{if $\frac{x_1 - \mu_1}{\sigma_1}$ \textgreater $\frac{x_2 - \mu_2}{\sigma_2}$}
\end{cases}
\end{align*}
where 
\begin{align*}
\frac{1}{p}f_1(x_1, x_2)&=\frac{(\alpha_0 + \alpha_1 + \alpha_2)}{(\alpha_1 + \alpha_2)}\frac{\alpha_1 (\alpha_0 + \alpha_2)}{\sigma_1 \sigma_2}\Big(1 + \frac{x_2 - \mu_2}{\sigma_2}\Big)^{-\alpha_0 - \alpha_2 - 1}\Big(1 + \frac{x_1 - \mu_1}{\sigma_1}\Big)^{-\alpha_1 - 1}\\
\frac{1}{p}f_2(x_1, x_2)&= \frac{(\alpha_0 + \alpha_1 + \alpha_2)}{(\alpha_1 + \alpha_2)}\frac{\alpha_2 (\alpha_0 + \alpha_1)}{\sigma_1 \sigma_2}\Big(1 + \frac{x_2 - \mu_2}{\sigma_2}\Big)^{-\alpha_2 - 1}\Big(1 + \frac{x_1 - \mu_1}{\sigma_1}\Big)^{-\alpha_0 - \alpha_1 - 1}
\end{align*}
We denote this distribution as $BBBVPA(\mu_1, \mu_2, \sigma_1, \sigma_2, \alpha_0, \alpha_1, \alpha_2)$. One can easily calculate the marginal distribution of $X_1$ and $X_2$. The pdf expressions are as follows:
\begin{align*}
f_{X_1}(x_1)&=\frac{\alpha_0 + \alpha_1 + \alpha_2}{\alpha_1 + \alpha_2}\Bigg\{\frac{\alpha_0 + \alpha_1}{\sigma_1}\Big(1 + \frac{x_1 - \mu_1}{\sigma_1}\Big)^{-\alpha_0 - \alpha_1 - 1} - \frac{\alpha_0}{\sigma_1}\Big(1 + \frac{x_1 - \mu_1}{\sigma_1}\Big)^{-\alpha_0 - \alpha_1 - \alpha_2 - 1}\Bigg\}\\
f_{X_2}(x_2)&=\frac{\alpha_0 + \alpha_1 + \alpha_2}{\alpha_1 + \alpha_2}\Bigg\{\frac{\alpha_0 + \alpha_2}{\sigma_2}\Big(1 + \frac{x_2 - \mu_2}{\sigma_2}\Big)^{-\alpha_0 - \alpha_2 - 1} - \frac{\alpha_0}{\sigma_2}\Big(1+\frac{x_2-\mu_2}{\sigma_2}\Big)^{-\alpha_0-\alpha_1-\alpha_2-1}\Bigg\}
\end{align*}

\subsection{Likelihood Function}

  Now we divide our data set $I = \{(x_{11}, x_{21}), (x_{12}, x_{22}) \cdots, (x_{1n}, x_{2n})\}$ into three parts.  They are as follow,
$ I_0 = \{i :\frac{x_{1i} - \mu_{1}}{\sigma_{1}} = \frac{x_{2i} - \mu_{2}}{\sigma_{2}} \}$, $ I_1 = \{i : \frac{x_{1i} - \mu_{1}}{\sigma_{1}} < \frac{x_{2i} - \mu_{2}}{\sigma_{2}} \}$, $ I_2 = \{ i : \frac{x_{1i} - \mu_{1}}{\sigma_{1}} > \frac{x_{2i} - \mu_{2}}{\sigma_{2}} \} $ and $n_0 = |I_0|$, $n_1 = |I_1|$, $n_2 = |I_2|$ where $|I_i|$ denotes number of observations in $I_{i}$.  Total number of observations $n = n_0 + n_1 + n_2.$

  Therefore log-likelihood function takes the form,
\begin{eqnarray*}
&& L(\mu_{1}, \mu_{2}, \sigma_{1}, \sigma_{2}, \alpha_{0}, \alpha_{1}, \alpha_{2})\\ & = & n\ln(\alpha_{0} + \alpha_{1} + \alpha_{2}) - n\ln(\alpha_{1} + \alpha_{2}) + n_{1}\ln\alpha_{1} + n_{1}\ln(\alpha_{0} + \alpha_{2})\\ & - & n_{1}\ln\sigma_{1} - n_{1}\ln\sigma_{2} - (\alpha_{0} + \alpha_{2} + 1)\sum_{i \in I_{1}}^{} \ln(1 + \frac{x_{2i} - \mu_{2}}{\sigma_{2}})\\ & - & (\alpha_{1} + 1)\sum_{i \in I_{1}}^{} \ln(1 + \frac{x_{1i} - \mu_{1}}{\sigma_{1}}) - n_{2}\ln\sigma_{1}  - n_{2}\ln\sigma_{2} + n_{2}\ln\alpha_{2}\\ & + & n_{2}\ln(\alpha_{0} + \alpha_{1}) - (\alpha_{0} + \alpha_{1} + 1)\sum_{i \in I_{2}}^{} \ln (1 + \frac{x_{1i} - \mu_{1}}{\sigma_{1}})\\ & - & (\alpha_{2} + 1)\sum_{i \in I_{2}}^{} \ln(1 + \frac{x_{2i} - \mu_{2}}{\sigma_{2}})
\end{eqnarray*}

\section{Bayesian Analysis}

 First step of all bayesian analysis follows bayesian mantra to calculate the posterior distribution.  We use $ posterior \propto likelihood \times prior$.  Therefore we need a suitable choice of set of priors in this case. 
  
\subsection{Prior Assumption}
\label{s4}

  We assume that the shape parameters $\alpha_0$, $\alpha_1$ and  $\alpha_2$ are distributed according to the gamma distribution with shape parameters $k_i$ and scale parameters $\theta_i$, i.e.,
\begin{align*}
\pi_0(\alpha_0) = \mathrm{Gamma}(k_0, \theta_0) = \frac{1}{\Gamma(k_0)\theta_0^{k_0}}\alpha_0^{k_0 - 1}e^{-\frac{\alpha_0}{\theta_0}}, \quad \alpha_0 > 0\\ 
\pi_1(\alpha_1) = \mathrm{Gamma}(k_1, \theta_1) = \frac{1}{\Gamma(k_1)\theta_1^{k_1}}\alpha_1^{k_1 - 1}e^{-\frac{\alpha_1}{\theta_1}}, \quad \alpha_1 > 0\\
\pi_2(\alpha_2) = \mathrm{Gamma}(k_2, \theta_2) = \frac{1}{\Gamma(k_2)\theta_2^{k_2}}\alpha_2^{k_2 - 1}e^{-\frac{\alpha_2}{\theta_2}}, \quad \alpha_2 > 0
\end{align*}
where $k_i > 0$ and $\theta_i > 0$, $i = 0, 1, 2$. Here $\Gamma(k)$ is the gamma function evaluated at $k$.  Suppose, $f_{N}$ is the probability density function of normal distribution.  Let's assume that the location parameters $\mu_1$ and $\mu_2$ are distributed according to the truncated normal distribution with mean = $\mu^{'}_i$ and variance = $\sigma^{'}_i$, $ i = 1, 2$,
\begin{align*}
\pi_3(\mu_1) = \frac{f_{N}(\mu_1; \mu^{'}_1, \sigma^{'}_1)}{\Phi(\min\{x_{1i}\}; \mu^{'}_1, \sigma^{'}_1)} \\
\pi_4(\mu_2) = \frac{f_{N}(\mu_2; \mu^{'}_2, \sigma^{'}_2)}{\Phi(\min\{x_{2i}\}; \mu^{'}_2, \sigma^{'}_2)}
\end{align*}
Further, We assume that the scale parameters $\sigma_1$ and $\sigma_2$ are distributed according to the gamma distribution with the hyper-parameters $c_1$, $c_2$, $d_1$ and $d_2$,
\begin{align*}
\pi_5(\sigma_1) = \mathrm{Gamma}(c_1, d_1) = \frac{1}{\Gamma(c_1)d_1^{c_1}}\sigma_1^{c_1 - 1}e^{-\frac{\sigma_1}{d_1}}, \quad \sigma_1 > 0\\
\pi_6(\sigma_2) = \mathrm{Gamma}(c_2, d_2) = \frac{1}{\Gamma(c_2)d_2^{c_2}}\sigma_2^{c_2 - 1}e^{-\frac{\sigma_2}{d_2}}, \quad \sigma_2 > 0
\end{align*}
where $c_i > 0$ and $d_i > 0$, i = 1, 2.

 One important assumption here is prior distributions of all parameters are independent of each other.

\subsection{Proposed Methodology}

 In this paper we obtain the Bayes estimates through slice cum gibbs sampler.  To form the Gibbs sampler first, we need to find out the expressions for conditional distribution of each parameter given the other parameters and the data.  We provide the expressions for logarithm of those conditional distributions.

  The log conditional posterior distributions of $\alpha_0$, $\alpha_1$, $\alpha_2$, $\mu_1$, $\mu_2$, $\sigma_1$, and $\sigma_2$ are given by,
\begin{eqnarray}
&& \ln(\pi(\alpha_0\mid \alpha_1, \alpha_2, \mu_1, \mu_2, \sigma_1, \sigma_2, x_1, x_2)) \propto n\ln(\alpha_0 + \alpha_1 + \alpha_2) + n_{1}\ln(\alpha_{0} + \alpha_{2}) \nonumber\\&- & (\alpha_{0} + \alpha_{2} + 1)\sum_{i \in I_{1}} \ln(1 + \frac{x_{2i} - \mu_{2}}{\sigma_{2}}) + n_{2}\ln(\alpha_{0} + \alpha_{1})\nonumber\\& - & (\alpha_{0} + \alpha_{1} + 1)\sum_{i \in I_{2}}^{} \ln (1 + \frac{x_{1i} - \mu_{1}}{\sigma_{1}}) + (k_0 - 1)\ln\alpha_0 - \frac{\alpha_0}{\theta_0} \label{cond-alpha0}
\end{eqnarray}
\begin{eqnarray}
&& \ln(\pi(\alpha_1\mid \alpha_0, \alpha_2, \mu_1, \mu_2, \sigma_1, \sigma_2, x_1, x_2)) \propto n\ln(\alpha_0 + \alpha_1 + \alpha_2) - n\ln(\alpha_1 + \alpha_2) + n_{1}\ln\alpha_{1}\nonumber\\& - & (\alpha_{1} + 1)\sum_{i \in I_{1}}^{} \ln(1 + \frac{x_{1i} - \mu_{1}}{\sigma_{1}}) + n_{2}\ln(\alpha_{0} + \alpha_{1})\nonumber\\& - & (\alpha_{0} + \alpha_{1} + 1)\sum_{i \in I_{2}}^{} \ln (1 + \frac{x_{1i} - \mu_{1}}{\sigma_{1}}) + (k_1-1)\ln\alpha_1-\frac{\alpha_1}{\theta_1}  \label{cond-alpha1}
\end{eqnarray}
\begin{eqnarray}
&& \ln(\pi(\alpha_2\mid \alpha_0, \alpha_1, \mu_1, \mu_2, \sigma_1, \sigma_2, x_1, x_2)) \propto n\ln(\alpha_0 + \alpha_1 + \alpha_2) - n\ln(\alpha_1 + \alpha_2)\nonumber\\ & + & n_{1}\ln(\alpha_{0} + \alpha_{2}) - (\alpha_{0} + \alpha_{2} + 1)\sum_{i \in I_{1}} \ln(1 + \frac{x_{2i} - \mu_{2}}{\sigma_{2}}) +  n_{2}\ln\alpha_{2}\nonumber\\ & - & (\alpha_{2} + 1)\sum_{i \in I_{2}}^{} \ln(1 + \frac{x_{2i} - \mu_{2}}{\sigma_{2}}) + (k_2-1)\ln\alpha_2-\frac{\alpha_2}{\theta_2}   \label{cond-alpha2}
\end{eqnarray}  
\begin{eqnarray}
&&\ln(\pi(\mu_1 \mid \alpha_0, \alpha_1, \alpha_2, \mu_2, \sigma_1, \sigma_2, x_1, x_2)) \propto -(\alpha_{1} + 1)\sum_{i \in I_{1}}^{} \ln(1 + \frac{x_{1i} - \mu_{1}}{\sigma_{1}})\nonumber\\ & - & (\alpha_{0} + \alpha_{1} + 1)\sum_{i \in I_{2}}^{} \ln (1 + \frac{x_{1i} - \mu_{1}}{\sigma_{1}}) - 0.5 \frac{(\mu_1- \mu'_1)^2}{(\sigma'_1)^{2}} \label{cond-mu1}
\end{eqnarray}
\begin{eqnarray}
&& \ln(\pi(\mu_2 \mid \alpha_0, \alpha_1, \alpha_2, \mu_1, \sigma_1, \sigma_2, x_1, x_2)) \propto - (\alpha_{0} + \alpha_{2} + 1)\sum_{i \in I_{1}}^{} \ln(1 + \frac{x_{2i} - \mu_{2}}{\sigma_{2}})\nonumber\\ & - & (\alpha_{2} + 1)\sum_{i \in I_{2}}^{} \ln(1 + \frac{x_{2i} - \mu_{2}}{\sigma_{2}}) - 0.5 \frac{(\mu_2 - \mu'_2)^2}{(\sigma'_2)^{2}}   \label{cond-mu2}
\end{eqnarray}
\begin{eqnarray}
&&\ln(\pi(\sigma_1\mid \alpha_0, \alpha_1, \alpha_2, \mu_1, \mu_2, \sigma_2, x_1, x_2)) \propto  -n_{1}\ln\sigma_{1} - (\alpha_{1} + 1)\sum_{i \in I_{1}}^{} \ln(1 + \frac{x_{1i} - \mu_{1}}{\sigma_{1}})\nonumber\\ & - & n_{2}\ln\sigma_{1} - (\alpha_{0} + \alpha_{1} + 1)\sum_{i \in I_{2}}^{} \ln (1 + \frac{x_{1i} - \mu_{1}}{\sigma_{1}}) + (c_1 - 1)\ln(\sigma_1) - \frac{\sigma_1}{d_1}  \label{cond-sigma1}
\end{eqnarray}
\begin{eqnarray}
&&\ln(\pi(\sigma_2\mid \alpha_0, \alpha_1, \alpha_2, \mu_1, \mu_2, \sigma_1, x_1, x_2)) \propto  - n_{1}\ln\sigma_{2} - n_{2}\ln\sigma_{2} \nonumber \\ & - & (\alpha_{0} + \alpha_{2} + 1)\sum_{i \in I_{1}}^{} \ln(1 + \frac{x_{2i} - \mu_{2}}{\sigma_{2}}) - (\alpha_{2} + 1)\sum_{i \in I_{2}}^{} \ln(1 + \frac{x_{2i} - \mu_{2}}{\sigma_{2}})\nonumber\\ & + & (c_2 - 1)\ln(\sigma_2) - \frac{\sigma_2}{d_2}  \label{cond-sigma2}
\end{eqnarray}

 We also require conditional distribution of $\mu_1$, $\mu_2$, $\sigma_1$ and $\sigma_2$ from posterior based on univariate likelihood function. 

\begin{eqnarray}
&&\ln(\pi(\mu_1 \mid \alpha_0, \alpha_1, \alpha_2, \mu_2, \sigma_1, \sigma_2, x_1)) \propto \sum_{i=1}^{n}\ln\Bigg\{\frac{\alpha_0+\alpha_1}{\sigma_1}\Big(1+\frac{x_{1i}-\mu_1}{\sigma_1}\Big)^{-\alpha_0-\alpha_1-1}\nonumber\\& - & \frac{\alpha_0}{\sigma_1}\Big(1+\frac{x_{1i}-\mu_1}{\sigma_1}\Big)^{-\alpha_0-\alpha_1-\alpha_2-1}\Bigg\} - 0.5 \frac{(\mu_1- \mu'_1)^2}{(\sigma'_1)^{2}} \label{cond-uni-mu1}
\end{eqnarray}

\begin{eqnarray}
&& \ln(\pi(\mu_2 \mid \alpha_0, \alpha_1, \alpha_2, \mu_1, \sigma_1, \sigma_2, x_2)) = \sum_{i=1}^{n}\ln\Bigg\{\frac{\alpha_0 + \alpha_2}{\sigma_2}\Big(1 + \frac{x_{2i} - \mu_2}{\sigma_2}\Big)^{- \alpha_0 - \alpha_2 - 1}\nonumber\\& - & \frac{\alpha_0}{\sigma_2}\Big(1 + \frac{x_{2i} - \mu_2}{\sigma_2}\Big)^{- \alpha_0 - \alpha_1 - \alpha_2 - 1}\Bigg\} - 0.5 \frac{(\mu_2 - \mu'_2)^2}{(\sigma'_2)^{2}} \label{cond-uni-mu2}
\end{eqnarray}

\begin{eqnarray}
&&\ln(\pi(\sigma_1\mid \alpha_0, \alpha_1, \alpha_2, \mu_1, \mu_2, \sigma_2, x_1)) \propto \sum_{i=1}^{n}\ln\Bigg\{\frac{\alpha_0+\alpha_1}{\sigma_1}\Big(1+\frac{x_{1i}-\mu_1}{\sigma_1}\Big)^{-\alpha_0-\alpha_1-1}\nonumber\\& - & \frac{\alpha_0}{\sigma_1}\Big(1+\frac{x_{1i}-\mu_1}{\sigma_1}\Big)^{-\alpha_0-\alpha_1-\alpha_2-1}\Bigg\} + (c_1 - 1)\ln(\sigma_1) - \frac{\sigma_1}{d_1}  \label{cond-uni-sigma1}
\end{eqnarray}

\begin{eqnarray}
&&\ln(\pi(\sigma_2\mid \alpha_0, \alpha_1, \alpha_2, \mu_1, \mu_2, \sigma_1, x_2)) \propto \sum_{i=1}^{n}\ln\Bigg\{\frac{\alpha_0+\alpha_2}{\sigma_2}\Big(1+\frac{x_{2i}-\mu_2}{\sigma_2}\Big)^{-\alpha_0-\alpha_2-1}\nonumber\\ & - & \frac{\alpha_0}{\sigma_2}\Big(1+\frac{x_{2i}-\mu_2}{\sigma_2}\Big)^{-\alpha_0-\alpha_1-\alpha_2-1}\Bigg\} + (c_2 - 1)\ln(\sigma_2) - \frac{\sigma_2}{d_2}  \label{cond-uni-sigma2}
\end{eqnarray}

  We propose to perform two variations of the algorithms to solve the problem using stepout method in slice sampling (\cite{Neal:2003}) to generate the respective parameters.  

\noindent{\textbf{Approach 1 :}} In this approach we propose to use usual step out methods for slice sampling for each conditional distributions. Standard slice sampling works only for continuous posterior. The algorithm is as follows :
\begin{itemize}
\item Sample z and u uniformly from the area under the distribution, say $p(\cdot)$.
\begin{enumerate}
\item Fix z, sample u uniformly from $[0, p(z)]$.
\item Fix u, sample z uniformly from the slice through the region $\{z : p(z) > u \}$
\end{enumerate}
\item How to sample z from the slice.
\begin{enumerate}
\item Start with the region of width w containing $z^{(t)}$.
\item If end point in slice, then extend region by w in that direction.
\item Sample $z^{'}$ uniformly from the region.
\item If $z^{'}$ is in the slice, the accept it as $z^{(t + 1)}$.
\item If not :  make $z^{'}$ new end point of the region, and resample $z^{'}$.
\end{enumerate}
\end{itemize} 
  
 Since the above algorithm needs continuous posterior, it is not going to work in this case.  Therefore we suggest the following modifications.
Instead of drawing sample from (\ref{cond-mu1}), (\ref{cond-mu2}), (\ref{cond-sigma1}), (\ref{cond-sigma2}),  we draw $\mu_1$, $\mu_2$, 
$\sigma_1$ and $\sigma_2$ from (\ref{cond-uni-mu1}), (\ref{cond-uni-mu2}), (\ref{cond-uni-sigma1}), (\ref{cond-uni-sigma2}) which are based on univariate likelihood.  
Therefore the steps of sampling are as follows :

\begin{itemize}
\item Start with some initial choice of parameters $\mu^{(0)}_{1}$, $\mu^{(0)}_{2}$, $\sigma^{(0)}_{1}$, $\sigma^{(0)}_{2}$, $\alpha^{(0)}_{0}$, $\alpha^{(0)}_{1}$ and $\alpha^{(0)}_{2}$.
\item Sample $\alpha^{(t + 1)}_{0}$, $\alpha^{(t + 1)}_{1}$ and $\alpha^{(t + 1)}_{2}$ from (\ref{cond-alpha0}), (\ref{cond-alpha1}) and (\ref{cond-alpha2}) using standard step-out slice sampling.
\item Sample $\mu^{(0)}_{1}$, $\mu^{(0)}_{2}$, $\sigma^{(0)}_{1}$, $\sigma^{(0)}_{2}$ from (\ref{cond-uni-mu1}), (\ref{cond-uni-mu2}), (\ref{cond-uni-sigma1}), (\ref{cond-uni-sigma2}) using standard step-out slice sampling. 
\item Go back to step-2.
\end{itemize}

\noindent{\textbf{Approach 2 :}}  In this approach we use directly discontinuous conditional posterior based on likelihood of bivariate distribution.  Using slice sampling on each of these conditional distributions from (\ref{cond-alpha0}) - (\ref{cond-sigma2}) is not straight forward.  We modify our slice sampling to fit for a discontinuous set up. Modified slice sampling steps are as follows :    

\begin{itemize}
\item Sample z and u uniformly from the area under the distribution, say $p(\cdot)$.
\begin{enumerate}
\item Fix z, sample u uniformly from $[0, p(z)]$.
\item Fix u, sample z uniformly from the slice through the region $\{z : p(z) > u \}$
\end{enumerate}
\item How to sample z from the slice.
\begin{enumerate}
\item Start with the region of width w containing $z^{(t)}$ \textbf{where it is defined}.
\item If end point in slice and \textbf{in the region where the function is defined}, then extend region by w in that direction.
\item If end point in slice and \textbf{in the region where the function is not defined}, then also extend region by w in that direction.
\item Sample $z^{'}$ uniformly from the region \textbf{until the point is defined}.
\item If $z^{'}$ is in the slice, the accept it as $z^{(t + 1)}$.
\item If not :  make $z^{'}$ new end point of the region, and resample $z^{'}$.
\item If resample point is not defined, \textbf{perform resampling until it is defined}.
\end{enumerate}
\end{itemize} 
Therefore the steps of sampling are as follows :
\begin{itemize}
\item Start with some initial choice of parameters $\mu^{(0)}_{1}$, $\mu^{(0)}_{2}$, $\sigma^{(0)}_{1}$, $\sigma^{(0)}_{2}$, $\alpha^{(0)}_{0}$, $\alpha^{(0)}_{1}$ and $\alpha^{(0)}_{2}$.
\item Sample $\alpha^{(t + 1)}_{0}$, $\alpha^{(t + 1)}_{1}$, $\alpha^{(t + 1)}_{2}$, $\mu^{(t + 1)}_{1}$, $\mu^{(t + 1)}_{2}$, $\sigma^{(t + 1)}_{1}$, $\sigma^{(t + 1)}_{2}$ from (\ref{cond-alpha0}), (\ref{cond-alpha1}), (\ref{cond-alpha2}), (\ref{cond-mu1}), (\ref{cond-mu2}), (\ref{cond-sigma1}) and (\ref{cond-sigma2}) respectively using modified slice sampling. 
\end{itemize}

\section{Constructing credible intervals for $\underline{\theta}$}

 We find the credible intervals for parameters as described by \cite{ChenShao:1999}.  Let assume $\underline{\theta}$ is vector.  To obtain credible intervals of first variable $\theta_{1i}$, we order $\{\theta_{1i}\}$, as $\theta_{1(1)} < \theta_{1(2)} < \cdots  < \theta_{1(M)}$. Then 100(1 - $\gamma$)$\%$ credible interval of $\theta_1$ become
$$(\theta_{1(j)}, \theta_{1(j + M - M\gamma)}), \qquad for \quad j = 1, \cdots , M\gamma$$

Therefore 100(1 - $\gamma$)$\%$ credible interval for $\theta_1$ becomes $(\theta_{1(j^*)}, \theta_{1(j^* + M - M\gamma)}),$ where $j^*$ is such that $$\theta_{1(j^* + M - M\gamma)} - \theta_{1(j^*)} \leq \theta_{1(j + M - M\gamma)} - \theta_{1(j)}$$
for all $j = 1, \cdots , M\gamma$. Similarly, we can obtain the credible interval for other co-ordinates of $\theta$.

 In this context we consider $\underline{\theta} = (\mu_1, \mu_2, \sigma_1, \sigma_2, \alpha_0, \alpha_1, \alpha_2)$ and $\gamma = 0.05$ to construct credible interval for all parameters.
      
\section{Numerical Result}  The numerical results are obtained by using package R 3.2.3.  The codes are run at IIT Guwahati computers with model : Intel(R) Core(TM) i5-6200U CPU 2.30 GHz. The codes will be available on request to authors. Bayes estimates, mean square errors, credible intervals are calculated by two approaches for two sets of parameters $\mu_1 = 0.3$, $\mu_2 = 0.4$, $\sigma_1 = 0.6$, $\sigma_2 = 0.7$, $\alpha_0 = 1.7$, $\alpha_1 = 1.2$, $\alpha_2 = 1.4$ and $\mu_1 = 1.0$, $\mu_2 = 2.0$, $\sigma_1 = 0.4$, $\sigma_2 = 0.5$, $\alpha_0 = 0.5$, $\alpha_1 = 0.70$, $\alpha_2 = 0.65$.  We use the following hyper parameters of prior as defined in Section 3.1 : $c_{1} = 0.1, d_{1} = 0.25, c_{2} = 3, d_{2} = 2, k_0 = 2, \theta_0 = 3, k_1 = 4, \theta_1 = 3, k_2 = 3, \theta_2 = 2, \mu^{'}_{1} = 0, \sigma^{'}_{1} = 1, \mu^{'}_{2} = 0, \sigma^{'}_{2} = 1$. However we observe that results does not vary much with different choice of hyper parameters. 

  Table-\ref{tab-absbvpa1}, Table-\ref{tab-absbvpa2}, Table-\ref{tab-absbvpa3} and Table-\ref{tab-absbvpa4} show bayes estimates, 95\% credible intervals for different parameter sets.  We also calculate mean square errors and coverage probabilities for 95\% credible intervals based on 200 different simulated samples from BB-BVPA.  Table-\ref{tab-absbvpa1}, Table-\ref{tab-absbvpa2} are results in Approach-1 whereas Table-\ref{tab-absbvpa3}, Table-\ref{tab-absbvpa4} are results obtained through Approach-2.  In slice cum gibbs sampling we take burn in period as 500.  Bayes estimates are calculated based on 2000 and more iterations after burn-in period.  We take bivariate sample size 450 and 1000 to get the estimates of the parameters. In both the slice sampling by step-out method, width is taken 1. However sampling procedure does not depend much on choice of width.  Numerical result show that estimates are very close to the original parameters.  Mean square errors are on higher side for shape parameters as compared to other parameters. However mean square errors decrease as sample size increases.  Coverage Probabilities are more closer to 95\% in Approach 2 than Approach 1.  This shows credible interval construction seems better fit for Approach 2 than Approach 1.  

\section{Case Study in real-life Data :}  We study one particular data sets which is used by \cite{DeyKundu:2017}.  The data set is taken from {\em{https://archive.ics.uci.edu/ml/machine-learning-databases}}.  The age of abalone is determined by cutting the shell through the cone, staining it, and counting the number of rings through a microscope. The data set contains related measurements.  We extract a part of the data for bivariate modeling.  We consider only measurements related to female population where one of the variable is Length as Longest shell measurement and other variable is Diameter which is perpendicular to length. We use peak over threshold method on this data set.  

  From \cite{FalkGuillou:2008}, we know that peak over threshold method on random variable U provides polynomial generalized Pareto distribution for any $x_{0}$ with $1 + log(G(x_{0})) \in (0, 1)$ i.e.  $P(U > t x_{0} | U > x_{0}) = t^{-\alpha}, ~~t \geq 1$ where $G(\cdot)$ is the distribution function of $U$.  We choose appropriate $t$ and $x_{0}$ so that data should behave more like Pareto distribution. The transformed data set does not have any singular component.  Therefore one possible assumption can be absolute continuous Marshall Olkin bivariate Pareto.

\begin{figure}[H]
 \begin{center}
  \subfigure[$\xi_{1}$]{\includegraphics[height=2.5in]{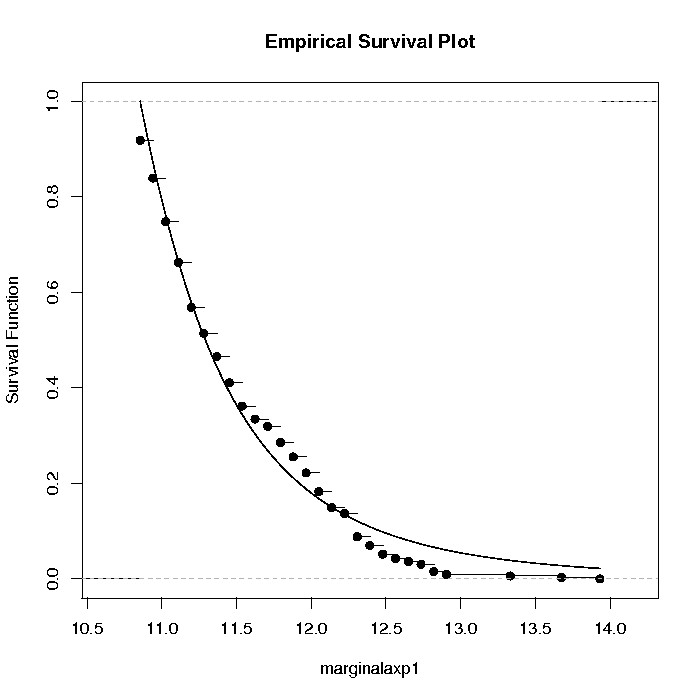}}
  \subfigure[$\xi_{2}$]{\includegraphics[height=2.5in]{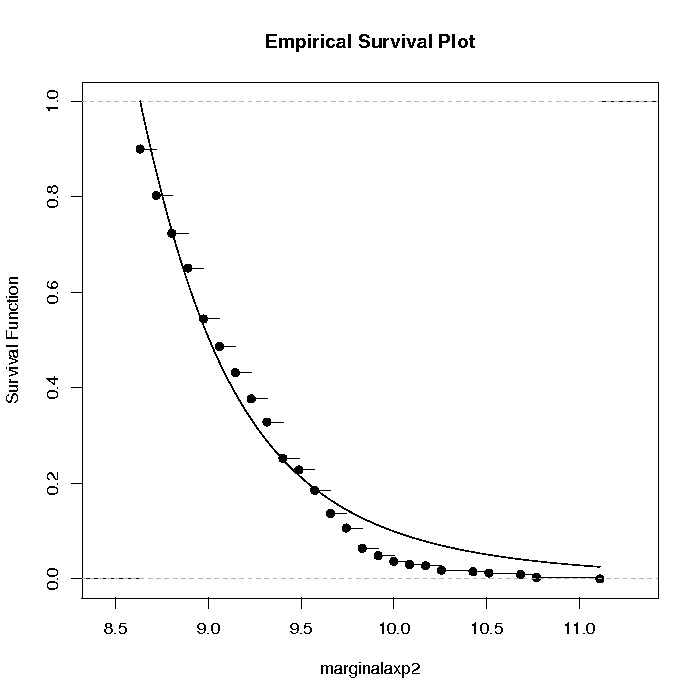}}\\
\caption{Survival plots for two marginals of the transformed dataset \label{figsurmarginalabalone}}
\end{center}
\end{figure}

  These data set are used to model seven parameter BB-BVPA.  EM estimates for BB-BVPA produces the values as $\mu_1 = 10.855$, $\mu_2 = 8.632$, $\sigma_1 = 2.124$, $\sigma_2 = 1.7110$, $\alpha_0 = 3.124$, $\alpha_1 = 1.743$, $\alpha_2 = 1.602$.  Figure-\ref{figsurmarginalabalone} shows that the empirical marginals coincide with the marginals calculated from the estimated parameters.  

\begin{figure}[H]	
 \begin{center}
  \includegraphics[width = 0.45\textwidth]{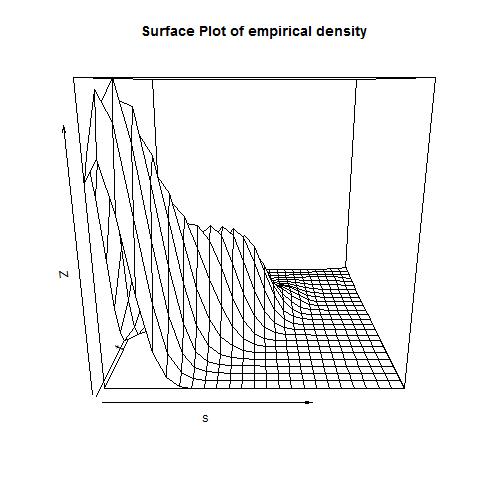} 
\caption{Two dimensional density plots for the transformed dataset \label{figdensurfaceabalone}}
\end{center}
\end{figure}

  We also verify our assumption by plotting empirical two dimensional density plot in Figure-\ref{figdensurfaceabalone} which resembles closer to the surface of absolute continuous Marshall-Olkin bivariate Pareto distribution. 
     
 We observe number of singular observations after transformation is zero.  Therefore it is reasonable to model the data with absolute continuous Marshall Olkin bivariate Pareto distribution.  Sample size for the data set is 329.  Note that even after location scale transformation observed cardinalities of $I_{1}$ and $I_{2}$ are good representative for the actual ones.  Bayesian estimates are calculated and provided in Table-\ref{Approach1:tab-data-est} and Table-\ref{Approach2:tab-data-est} using Approach 1 and Approach 2 respectively for the abalone data.  We use parametric bootstrap technique to generate different samples from seven parameter BB-BVPA using estimated parameters as original parameter and then find out the mean square error and coverage probabilities for the constructed credible interval.  Results are available for in Table-\ref{tab-dataana-app1} and Table-\ref{tab-dataana-app2} using Approach 1 and Approach 2 respectively. Both the algorithm works quite well in estimating all parameters. However we observe that the mean square error is on higher side for $\alpha_0$ as compared to other parameters.  It is also unable to capture the coverage probability of $\alpha_0$ at desired level and provides a value significantly smaller than the desired 95\% confidence level. However we get significant improvement in both MSEs and Coverage Probabilities if we increase the sample size.  Therefore it is safer to work with larger sample size to obtain a desired accuracy level while applying these methods in real life data sets. 

\section{Conclusion}  Bayes estimates of the parameters of absolute continuous bivariate Pareto under square error loss are obtained using Slice cum Gibbs sampler approach.  Two approaches proposed in this paper works quite well even for moderately large sample size.  However coverage probabilities calculated based on parametric bootstrap samples show that the proposed credible interval construction does not work very well for all the shape parameters.  Although we see its significant improvement for large sample, more research is needed to explore the best credible interval construction in small sample set up.  The same study can be made using many other algorithms like importance sampling, HMC etc. This estimation procedure can be used in bayesian discrimination between two bivariate distributions using bayes factor. The work is on progress.  

\nocite{SarhanBalakrishnan:2007} \nocite{RisticKundu:2015} \nocite{RakonczaiZempleni:2012} \nocite{MirhosseiniAminiKunduDolati:2015} \nocite{MarshallOlkin:1967} \nocite{KunduKumarGupta:2015} \nocite{KunduGupta:2010} \nocite{KhosraviKunduJamalizadeh:2015} \nocite{KunduGupta:2011} \nocite{Kundu:2012} \nocite{Hanagal:1996} \nocite{FalkGuillou:2008} \nocite{ChristianCasella:1999} \nocite{Arnold:1967}

\bibliographystyle{apa}

\bibliography{bvpa_bayes}

\begin{table}[ht!]
{\begin{tabular}[l]{@{}lcccccccc}\hline
Slice-cum-Gibbs & & & & & & &\\ \hline
 Gamma Prior & & & & & & & \\ \hline
 n = 329 & & & & & & &\\ \hline
Original Parameter Sets & $\mu_1 $ & $\mu_2$ & $\sigma_1$ & $\sigma_2$ & $\alpha_0$ & $\alpha_1$ & $\alpha_2$ \\
 Starting Value & 5.204 & 7.470 & 0.778 & 0.539 & 0.442 & 0.614 & 0.455 \\  
  Bayes Estimates & 10.853 & 8.630 & 1.365 & 1.188 & 2.020 & 1.245 & 1.218 \\  \hline
\end{tabular}}   
\caption{The Bayes Estimates of the parameters of absolute continuous Marshall-Olkin bivariate Pareto distribution based on the abalone data set}
\label{Approach1:tab-data-est} 
\end{table}

\begin{table}[ht!]
{\begin{tabular}[l]{@{}lcccccccc}\hline
Slice-cum-Gibbs & & & & & & &\\ \hline
 Gamma Prior & & & & & & & \\ \hline
 n = 329 & & & & & & &\\ \hline
Original Parameter Sets & $\mu_1 $ & $\mu_2$ & $\sigma_1$ & $\sigma_2$ & $\alpha_0$ & $\alpha_1$ & $\alpha_2$ \\
 Starting Value & 5.204 & 7.470 & 0.778 & 0.539 & 0.442 & 0.614 & 0.455 \\  
  Bayes Estimates & 10.853 & 8.630 & 2.316 & 1.923 & 4.049 & 1.240 & 1.158\\  \hline
\end{tabular}}   
\caption{The Bayes Estimates of the parameters of absolute continuous Marshall-Olkin bivariate Pareto distribution based on the abalone data set}
\label{Approach2:tab-data-est} 
\end{table}

\begin{table}[H]
{\begin{tabular}[l]{@{}lccccc}\hline
Slice-cum-Gibbs & & & &\\ \hline
 Gamma Prior & & & & \\ \hline
 n = 329 & & & & \\ \hline
 Parameter Sets & $\mu_1$ & $\mu_2$ & $\sigma_1$ & $\sigma_2$ \\ 
 Starting Value & 5.203 & 7.469 & 0.778 & 0.539 \\     
 Average Bayes Estimates & 10.852 & 8.6303 & 1.365 & 1.188 \\ 
  Mean Square Error & 4.374$\times$ $10^{-6}$ & 3.598 $\times$ $10^{-6}$ & 0.660 & 0.371 \\              
  Credible Intervals & [10.854, 10.859] & [8.628, 8.632] & [0.699, 1.674] & [0.691, 1.492] \\ 
  Coverage Probability & 0.965 & 0.92 & 0.48 & 0.59 \\ \hline 
 n = 329 & & & & \\ \hline
 Parameter Sets & & $\alpha_0$ & $\alpha_1$ & $\alpha_2$ \\ 
 Starting Value & & 0.442 & 0.614 & 0.455 \\     
 Average Bayes Estimates & & 2.020 & 1.245 & 1.177 \\ 
  Mean Square Error &  & 2.198 & 0.046 & 0.043 \\              
  Credible Intervals & & [0.804, 2.866] & [0.029, 1.362] & [0.284, 1.538] \\ 
  Coverage Probability & & 0.43 & 1 & 1 \\ \hline   
 n = 1000 & & & & \\ \hline
 Parameter Sets & $\mu_1$ & $\mu_2$ & $\sigma_1$ & $\sigma_2$ \\ 
 Starting Value & 5.203 & 7.469 & 0.778 & 0.539 \\      
 Average Bayes Estimates & 10.853 & 8.630 & 1.605 & 1.295 \\ 
  Mean Square Error & 3.589 $\times$ $10^{-07}$ & 4.556 $\times$ $10^{-07}$ & 0.082 & 0.033 \\              
  Credible Intervals & [10.851, 10.853] & [8.630, 8.632] & [0.815, 1.443] & [0.861, 1.338]\\ 
  Coverage Probability & 0.96 & 0.94 & 0.90 & 0.93 \\ \hline
 n = 1000 & & & & \\ \hline
 Parameter Sets & & $\alpha_0$ & $\alpha_1$ & $\alpha_2$ \\ 
 Starting Value & & 0.442 & 0.614 & 0.455 \\      
 Average Bayes Estimates & & 2.829 & 1.172 & 1.064\\ 
  Mean Square Error & & 0.374 & 0.199 & 0.087 \\              
  Credible Intervals & & [1.849, 2.793] & [0.330, 1.003] & [0.337, 0.986]\\ 
  Coverage Probability & & 0.78 & 0.965 & 0.99 \\ \hline
\end{tabular}}   
\caption{Results for Approach 1 : The Bayes Estimates (BE), Mean Square Error (MSE) and credible intervals of absolute continuous Marshall-Olkin bivariate Pareto distribution with parameters $\mu_1 = 10.852$, $\mu_2 = 8.630$, $\sigma_1 = 2.316$, $\sigma_2 = 1.923$, $\alpha_0 = 4.049$, $\alpha_1 = 1.240$, $\alpha_2 = 1.158$.}
\label{tab-dataana-app1} 
\end{table}

\begin{table}[H]
{\begin{tabular}[l]{@{}lccccc}\hline
Slice-cum-Gibbs & & & &\\ \hline
 Gamma Prior & & & & \\ \hline
 n = 329 & & & & \\ \hline
 Parameter Sets & $\mu_1$ & $\mu_2$ & $\sigma_1$ & $\sigma_2$ \\ 
 Starting Value & 5.203 & 7.469 & 0.778 & 0.539 \\     
 Average Bayes Estimates & 10.852 & 8.6303 & 1.519 & 1.330 \\ 
  Mean Square Error & 4.374$\times$ $10^{-6}$ & 3.598 $\times$ $10^{-6}$ & 0.660 & 0.371 \\              
  Credible Intervals & [10.854, 10.859] & [8.628, 8.632] & [0.699, 1.674] & [0.691, 1.492] \\ 
  Coverage Probability & 0.965 & 0.92 & 0.48 & 0.59 \\ \hline 
n = 329 & & & & \\ \hline
 Parameter Sets & & $\alpha_0$ & $\alpha_1$ & $\alpha_2$ \\ 
 Starting Value & & 0.442 & 0.614 & 0.455 \\     
 Average Bayes Estimates & & 2.871 & 1.216 & 1.177 \\ 
  Mean Square Error &  & 2.198 & 0.046 & 0.043 \\              
  Credible Intervals & & [0.804, 2.866] & [0.029, 1.362] & [0.284, 1.538] \\ 
  Coverage Probability & & 0.43 & 1 & 1 \\ \hline   
 n = 1000 & & & & \\ \hline
 Parameter Sets & $\mu_1$ & $\mu_2$ & $\sigma_1$ & $\sigma_2$ \\ 
 Starting Value & 5.203 & 7.469 & 0.778 & 0.539 \\      
 Average Bayes Estimates & 10.853 & 8.630 & 1.849 & 1.608 \\ 
  Mean Square Error & 6.156$\times$ $10^{-7}$ & 1.86$\times$ $10^{-6}$ & 0.549 & 0.289 \\              
  Credible Intervals & [10.852, 10.854] & [8.629, 8.630] & [1.230, 2.077] & [1.054, 1.763]\\ 
  Coverage Probability & 0.91 & 0.955 & 0.775 & 0.88 \\ \hline
 n = 1000 & & & & \\ \hline
 Parameter Sets & & $\alpha_0$ & $\alpha_1$ & $\alpha_2$ \\ 
 Starting Value & & 0.442 & 0.614 & 0.455 \\      
 Average Bayes Estimates & & 3.555 & 1.145 & 1.124 \\ 
  Mean Square Error & & 1.863 & 0.049 & 0.041 \\              
  Credible Intervals & & [1.918, 3.655] & [0.601, 1.712] & [0.598, 1.599]\\ 
  Coverage Probability & & 0.77 & 1 & 1 \\ \hline
\end{tabular}}   
\caption{Results for Approach 2 : The Bayes Estimates (BE), Mean Square Error (MSE) and credible intervals of absolute continuous Marshall-Olkin bivariate Pareto distribution with parameters $\mu_1 = 10.852$, $\mu_2 = 8.630$, $\sigma_1 = 2.316$, $\sigma_2 = 1.923$, $\alpha_0 = 4.049$, $\alpha_1 = 1.240$, $\alpha_2 = 1.158$.}
\label{tab-dataana-app2} 
\end{table}

\begin{table}[ht!]
\begin{small}
{\begin{tabular}[l]{@{}lccccc}\hline
Slice-cum-Gibbs & & & &\\ \hline
 Gamma Prior & & & & \\ \hline
 n = 450 & & & & \\ \hline
Original Parameter Sets & $\mu_1 = 0.3$ & $\mu_2 = 0.4$ & $\sigma_1 = 0.6$ & $\sigma_2 = 0.7$ \\ 
 Starting Value & 0.143 & 0.346 & 0.775 & 0.539 \\     
  Bayes Estimates & 0.300 & 0.400 & 0.490 & 0.627  \\ 
  Mean Square Error & 0.00000052 & 0.00000049 & 0.039 & 0.012 \\              
  Credible Intervals & [0.298, 0.301] & [0.398, 0.401] & [0.269, 0.592] & [0.149, 0.752] \\ 
  Coverage Probability & 0.93 & 0.935 & 0.98 & 0.89 \\ \hline
 n = 450 & & & & \\ \hline
 Original Parameter Sets & & $\alpha_0 = 1.7$ & $\alpha_1 = 1.2$ & $\alpha_2 = 1.4$ \\ 
 Starting Value && 0.442 & 0.614 & 0.455 \\     
  Bayes Estimates && 1.526 & 0.889 & 1.095 \\ 
  Mean Square Error && 0.292 & 0.636 & 0.108 \\              
  Credible Intervals & & [0.150, 0.752] & [0.3298, 1.058] & [0.526, 1.426]\\ 
  Coverage Probability & & 0.715 & 0.945 & 0.99 \\ \hline
 n = 1000 & & & & \\ \hline
 Original Parameter Sets & $\mu_1 = 0.3$ & $\mu_2 = 0.4$ & $\sigma_1 = 0.6$ & $\sigma_2 = 0.7$ \\ 
 Starting Value & 0.144 & 0.346 & 0.778 & 0.539 \\     
  Bayes Estimates & 0.300 & 0.400 & 0.490 & 0.627  \\ 
  Mean Square Error & 0.00000052 & 0.00000062 & 0.029 & 0.024  \\              
  Credible Intervals & [0.298, 0.301] & [0.398, 0.400] & [0.312, 0.547] & [0.401, 0.629] \\ 
  Coverage Probability & 0.995 & 0.975 & 0.945 & 0.895 \\ \hline
 n = 1000 & & & & \\ \hline
 Original Parameter Sets & & $\alpha_0 = 1.7$ & $\alpha_1 = 1.2$ & $\alpha_2 = 1.4$ \\ 
 Starting Value & & 0.441 & 0.614 & 1.425 \\      
 Bayes Estimates & & 1.852 & 0.844 & 1.011\\ 
 Mean Square Error & &  0.058 & 0.213 & 0.254 \\              
 Credible Intervals & & [0.829, 1.446] & [0.645, 1.295] & [0.5501, 1.197] \\ 
 Coverage Probability & & 0.79 & 0.945 & 0.96 \\ \hline
\end{tabular}}   
\end{small}
\caption{Approach 1 : The Bayes Estimates (BE), Mean Square Error (MSE) and credible intervals of absolute continuous Marshall-Olkin bivariate Pareto distribution with parameters $\mu_1 = 0.3$, $\mu_2 = 0.4$, $\sigma_1 = 0.6$, $\sigma_2 = 0.7$, $\alpha_0 = 1.7$, $\alpha_1 = 1.2$, $\alpha_2 = 1.4$.}
\label{tab-absbvpa1} 
\end{table}

\begin{table}[ht!]
\begin{small}
{\begin{tabular}[l]{@{}lccccc}\hline
Slice-cum-Gibbs & & & &\\ \hline
 Gamma Prior & & & & \\ \hline
 n = 450 & & & & \\ \hline
 Original Parameter Sets & $\mu_1 = 1.0$ & $\mu_2 = 2.0$ & $\sigma_1 = 0.4$ & $\sigma_2 = 0.5$ \\ 
 Starting Value & 0.157 & 0.656 & 0.153 & 0.645  \\     
  Bayes Estimates & 0.999 & 1.1999 & 0.399 & 0.608 \\ 
  Mean Square Error & 6.233$\times$ $10^{-7}$ & 9.398$\times$ $10^{-7}$ & 0.011 & 0.023 \\              
  Credible Intervals & [0.998, 1.0006] & [1.9986, 2.0011] & [0.1684, 0.3552] & [0.2401, 0.5190] \\ 
  Coverage Probability & 0.96 & 0.96 & 0.96 & 0.94 \\ \hline  
n = 450 & & & & \\ \hline
 Original Parameter Sets & & $\alpha_0 = 0.50$ & $\alpha_1 = 0.70$ & $\alpha_2 = 0.65$ \\ 
 Starting Value & & 0.523 & 0.253 & 0.478 \\     
  Bayes Estimates & & 0.731 & 0.671 & 0.705\\ 
  Mean Square Error & & 0.063 & 0.098 & 0.1009 \\              
  Credible Intervals & & [0.463, 0.956] & [0.264, 0.714] & [0.305, 0.758]\\ 
  Coverage Probability & & 0.78 & 0.93 & 0.89 \\ \hline
 n = 1000 & & & & \\ \hline
 Original Parameter Sets & $\mu_1 = 1.0$ & $\mu_2 = 2.0$ & $\sigma_1 = 0.4$ & $\sigma_2 = 0.5$ \\ 
 Starting Value & 0.812 & 0.747 & 0.567 & 0.818  \\      
 Bayes Estimates & 1.000 & 2.000 & 0.445 & 0.609  \\ 
  Mean Square Error & 0.00000014 & 0.00000028 & 0.0084 & 0.0217  \\              
  Credible Intervals & [0.9992, 1.0003] & [1.999, 2.0006] & [0.2139, 0.3671] & [0.3366, 0.5136]  \\ 
  Coverage Probability & 0.95 & 0.945 & 0.92 & 0.965 \\ \hline
n = 1000 & & & & \\ \hline
 Original Parameter Sets & & $\alpha_0 = 0.50$ & $\alpha_1 = 0.70$ & $\alpha_2 = 0.65$ \\ 
 Starting Value &  & 0.387 & 0.0.227 & 0.059 \\      
 Bayes Estimates &  & 0.857 & 0.682 & 0.694 \\ 
  Mean Square Error &  & 0.0182 & 0.03004 & 0.0274 \\              
  Credible Intervals &  & [0.6762, 1.0229] & [0.3201, 0.6573] & [0.2906, 0.6196] \\ 
  Coverage Probability & & 0.895 & 0.915 & 0.925 \\ \hline
\end{tabular}}   
\end{small}
\caption{Approach 1 : The Bayes Estimates (BE), Mean Square Error (MSE) and credible intervals of absolute continuous Marshall-Olkin bivariate Pareto distribution with parameters $\mu_1 = 1.0$, $\mu_2 = 2.0$, $\sigma_1 = 0.4$, $\sigma_2 = 0.5$, $\alpha_0 = 0.5$, $\alpha_1 = 0.70$, $\alpha_2 = 0.65$.}
\label{tab-absbvpa2} 
\end{table}

\begin{table}[ht!]
\begin{small}
{\begin{tabular}[l]{@{}lccccc}\hline
Slice-cum-Gibbs & & & &\\ \hline
 Gamma Prior & & & & \\ \hline
 n = 450 & & & & \\ \hline
 Original Parameter Sets & $\mu_1 = 1.0$ & $\mu_2 = 2.0$ & $\sigma_1 = 0.4$ & $\sigma_2 = 0.5$ \\ 
 Starting Value & 0.479 & 1.731 & 0.778 & 0.539  \\     
  Bayes Estimates & 1.00004 & 2.00004 & 0.432 & 0.582 \\ 
  Mean Square Error & 6.6623$\times$ $10^{-7}$ & 1.265 $\times$ $10^{-7}$ & 0.0059 & 0.0155 \\              
  Credible Intervals & [0.998, 1.0005] & [1.998, 2.0006] & [0.209, 0.414] & [0.293, 0.539] \\ 
  Coverage Probability & 0.93 & 0.935 & 0.99 & 0.96 \\ \hline  
n = 450 & & & & \\ \hline
 Original Parameter Sets & & $\alpha_0 = 0.50$ & $\alpha_1 = 0.70$ & $\alpha_2 = 0.65$ \\ 
 Starting Value & & 0.442 & 0.614 & 0.455 \\     
  Bayes Estimates & & 0.603 & 0.903 & 0.888 \\ 
  Mean Square Error & & 0.090 & 0.098 & 0.1119 \\              
  Credible Intervals & & [0.00012, 0.3065] & [0.786, 1.353] & [0.377, 0.935]\\ 
  Coverage Probability & & 0.775 & 0.915 & 0.86 \\ \hline
 n = 1000 & & & & \\ \hline
 Original Parameter Sets & $\mu_1 = 1.0$ & $\mu_2 = 2.0$ & $\sigma_1 = 0.4$ & $\sigma_2 = 0.5$ \\ 
 Starting Value & 0.479 & 1.731 & 0.778 & 0.539  \\      
 Bayes Estimates & 1.000 & 1.999 & 0.405 & 0.544  \\ 
  Mean Square Error & 0.00000014 & 0.00000028 & 0.0018 & 0.0053 \\              
  Credible Intervals & [0.999, 1.00008] & [1.999, 2.0002] & [0.226, 0.364] & [0.321, 0.505]  \\ 
  Coverage Probability & 0.945 & 0.955 & 0.98 & 0.975 \\ \hline
n = 1000 & & & & \\ \hline
 Original Parameter Sets & & $\alpha_0 = 0.50$ & $\alpha_1 = 0.70$ & $\alpha_2 = 0.65$ \\ 
 Starting Value &  & 0.442 & 0.614 & 0.455 \\      
 Bayes Estimates &  & 0.729 & 0.761 & 0.748 \\ 
  Mean Square Error & & 0.021 & 0.021 & 0.026 \\              
  Credible Intervals &  & [0.504, 0.914] & [0.373, 0.792] & [0.278, 0.657] \\ 
  Coverage Probability & & 0.925 & 0.975 & 0.955 \\ \hline
\end{tabular}}   
\end{small}
\caption{Approach 2 : The Bayes Estimates (BE), Mean Square Error (MSE) and credible intervals of absolute continuous Marshall-Olkin bivariate Pareto distribution with parameters $\mu_1 = 1.0$, $\mu_2 = 2.0$, $\sigma_1 = 0.4$, $\sigma_2 = 0.5$, $\alpha_0 = 0.5$, $\alpha_1 = 0.70$, $\alpha_2 = 0.65$.}
\label{tab-absbvpa3} 
\end{table}

\begin{table}[ht!]
\begin{small}
{\begin{tabular}[l]{@{}lccccc}\hline
Slice-cum-Gibbs & & & &\\ \hline
 Gamma Prior & & & & \\ \hline
 n = 450 & & & & \\ \hline
 Original Parameter Sets & $\mu_1 = 0.3$ & $\mu_2 = 0.4$ & $\sigma_1 = 0.6$ & $\sigma_2 = 0.70$ \\ 
 Starting Value & 0.144 & 0.346 & 0.778 & 0.539 \\     
 Average Bayes Estimates & 0.3001 & 0.40003 & 0.6376 & 0.6456 \\ 
  Mean Square Error & 6.002$\times$ $10^{-7}$ & 5.22 $\times$ $10^{-7}$ & 0.0099 & 0.0106 \\              
  Credible Intervals & [0.298, 0.300] & [0.399, 0.4003] & [0.321, 0.616] & [0.330, 0.636] \\ 
  Coverage Probability & 0.92 & 0.935 & 0.995 & 0.935 \\ \hline  
n = 450 & & & & \\ \hline
 Original Parameter Sets & & $\alpha_0 = 1.7$ & $\alpha_1 = 1.2$ & $\alpha_2 = 1.4$ \\ 
 Starting Value & & 0.442 & 0.614 & 0.455 \\   
 Average Bayes Estimates & & 1.359 & 1.530 & 1.474 \\ 
  Mean Square Error & & 0.236 & 0.259 & 0.125 \\              
  Credible Intervals & & [0.041, 1.038] & [0.572, 1.691] & [0.362, 1.365]\\ 
  Coverage Probability & & 0.84 & 0.98 & 0.98 \\ \hline
 n = 1000 & & & & \\ \hline
 Original Parameter Sets & $\mu_1 = 0.3$ & $\mu_2 = 0.4$ & $\sigma_1 = 0.6$ & $\sigma_2 = 0.7$ \\ 
 Starting Value & 0.479 & 1.731 & 0.778 & 0.539  \\      
 Average Bayes Estimates & 0.3000 & 0.3999 & 0.0.623 & 0.639  \\ 
  Mean Square Error & 8.28$\times$ $10^{-8}$ & 9.78 $\times$ $10^{-8}$ & 0.0054 & 0.0084 \\              
  Credible Intervals & [0.299, 0.300] & [0.3996, 0.4003] & [0.348, 0.556] & [0.380, 0.612]\\ 
  Coverage Probability & 0.95 & 0.95 & 0.98 & 0.895 \\ \hline
n = 1000 & & & & \\ \hline
 Original Parameter Sets & & $\alpha_0 = 1.7$ & $\alpha_1 = 1.2$ & $\alpha_2 = 1.4$ \\ 
 Starting Value &  & 0.442 & 0.614 & 0.455 \\      
 Average Bayes Estimates &  & 1.593 & 1.319 & 1.313 \\ 
  Mean Square Error & & 0.0580 & 0.062 & 0.054 \\              
  Credible Intervals &  & [1.096, 1.874] & [0.556, 1.334] & [0.598, 1.334] \\ 
  Coverage Probability & & 0.975 & 0.99 & 0.955 \\ \hline
\end{tabular}}   
\end{small}
\caption{Approach 2 : The Bayes Estimates (BE), Mean Square Error (MSE) and credible intervals of absolute continuous Marshall-Olkin bivariate Pareto distribution with parameters $\mu_1 = 0.3$, $\mu_2 = 0.4$, $\sigma_1 = 0.6$, $\sigma_2 = 0.7$, $\alpha_0 = 1.7$, $\alpha_1 = 1.2$, $\alpha_2 = 1.4$.}
\label{tab-absbvpa4} 
\end{table}

\end{document}